# Superconducting properties of the pyrochlore oxide $Cd_2Re_2O_7$

Z. Hiroi and M. Hanawa

*Institute for Solid State Physics, University of Tokyo, Kashiwanoha, Kashiwa, Chiba 277-8581, Japan*

We report the superconducting properties of the pyrochlore oxide $Cd_2Re_2O_7$. The bulk superconducting transition temperature $T_c$ is about 1.0 K, and the upper critical field $H_{c2}$ determined by the measurement of specific heat under magnetic fields is 0.29 T. The superconducting coherence length is estimated to be 34 nm. Specific heat data measured on single crystals suggest that the superconducting gap of $Cd_2Re_2O_7$ is nodeless.

## 1. Introduction

Recently various transition-metal (TM) oxides which exist in the vicinity of a metal-insulator (MI) transition have been studied extensively, and many intriguing phenomena such as high-temperature superconductivity in cuprates and charge/orbital ordering in manganites or others have been found. In many cases they crystallize in the perovskite-related structures where TM-O octahedra are linked by their vertices with nearly linear TM-O-TM bonds.

There is another class of compounds forming a large family of TM oxides which crystallize in the pyrochlore structure (Fig. 1) with a chemical formula $A_2B_2O_7$ where the B represents TMs [1]. The space group of the ideal cubic pyrochlore structure is $Fd3m$, and there are eight molecules per unit cell. The structure is highly symmetrical with only one variable atomic coordination parameter; $x$ for the O1 site. The TM cations (B) are six coordinated by O1 and are located within a rhombohedrally distorted octahedron (a trigonal antiprism). The above parameter $x$ of O1 decides the shape of a $BO_6$ octahedron: It becomes undistorted only when $x = 0.3125$, while compressed along the <111> direction for $x > 0.3125$, which is the general case for most pyrochlore oxides [1]. In contrast, it is believed that cadmium pyrochlores are exceptional to have $x \leq 0.3125$: $x = 0.309$ was reported for $Cd_2Re_2O_7$ [2].

Unlike the perovskite structure, the pyrochlore structure contains TM-O octahedra connected with significantly bent TM-O-TM bonds (110° ~ 140°). This structural difference in the TM-O network gives rise to a substantial difference in electronic structures between the two families. For example, a metallic ground state is generally more stable for perovskite oxides than for pyrochlore oxides. Moreover, when $d$ electrons are localized, an antiferromagnetic insulator is attained for the former, while a ferromagnetic one is favorable for the latter. Looking in the chemical trend of electronic properties for pyrochlore compounds, most $3d$ and $4d$ TM pyrochlores are insulators owing to large electron correlations as well as relatively small electron transfers along the bent B-O-B bonds. Molybdenum pyrochlores exist near the metal-insulator boundary, where ferromagnetic metals appear as the ionic radius of the counter cations is increased [3]. On one hand, metallic pyrochlores can be found, when additional electrons are supplied from A cations like typically in Mn and Ru pyrochlores such as $Tl_2Mn_2O_7$ and $Bi_2Ru_2O_7$ [4, 5]. In contrast, $5d$ TM pyrochlores are often metallic because of relatively spreading $5d$ orbitals [1]. A rare exception reported previously is found in $Os^{5+}$ pyrochlores like $Cd_2Os_2O_7$ [6, 7] and $Ca_2Os_2O_7$ [8], where a MI transition occurs with temperature. An $Os^{5+}$ ion has a $5d^3$ electron configuration and thus the $t_{2g}$ orbital is half-filled, suggesting a possible Mott-Hubbard type MI transition. Note that most of $Os^{4+}$ pyrochlores are metals. These examples illustrate that the effect of electron correlations is still important even for the $5d$ electron systems in the pyrochlore structure. No superconductivity has been observed there so far in spite of many metallic compounds present in the family. Very recently, we have searched for novel phenomena in pyrochlore compounds near $Cd_2Os_2O_7$ and found that $Cd_2Re_2O_7$ becomes the first superconductor in the pyrochlore family [9]. Sakai *et al.* also reported superconductivity in the same compound independently [10], and Jin *et al.* reported similar results later[11].

Since $Cd^{2+}$ and $Re^{5+}$ have $4d^{10}$ and $4f^{14}5d^2$ outer electron configurations, respectively, only $Re^{5+}$ is expected to underlie the electronic and magnetic properties of $Cd_2Re_2O_7$. In a simple ionic model, $Cd_2Re_2O_7$ possesses two $d$ electrons in the $t_{2g}$ orbitals which should form a band in a crystal, as schematically depicted in Fig. 1. According to the recent band structure calculations [12, 13], the band width is about 2.5 eV, and the contribution of oxygen $2p$ orbitals is small compared with the case of perovskite oxides. In the previous study on $Cd_2Re_2O_7$ Donohue *et al.* prepared a single crystal, determined the crystal structure, and reported the resistivity which was metallic above 4 K [2]. Blacklock and White measured the specific heat above 1.8 K using a polycrystalline sample and found the Sommerfeld coefficient $\gamma$ to be 13.3 mJ/K$^2$ mol Re [14]. We prepared single crystals of $Cd_2Re_2O_7$ and measured resistivity and specific heat down to ~0.4 K [9]. A sharp drop in resistivity and a large jump in specific heat were observed at $T = 1$ K, which give strong evidence for the occurrence of superconductivity. The upper critical field $H_{c2}$ was determined by the measurement of specific heat under magnetic fields. Various superconducting parameters were deduced.

## 2. Experimental

Single crystals of $Cd_2Re_2O_7$ were prepared by a chemical reaction, $2CdO + 5/3ReO_3 + 1/3Re \rightarrow Cd_2Re_2O_7$. Stoichiometric amounts of these starting powders were mixed in an agate mortar and pressed into a pellet. The reaction was carried out in an evaluated quartz ampoule at 800-900ºC for 70 h. The products appeared as purple octahedral crystals of a few mm on an edge which adhered to the walls of the ampoule. Several batches of crystals were prepared. The chemical composition of the crystals examined by the electron-probe microanalysis or the inductively coupled plasma analysis was Cd/Re = 1.00 ± 0.01. The oxygen content determined by the Iodometry was 7.10. A powder X-ray diffraction (XRD) pattern was taken at room temperature and indexed on the basis of a face-centered cubic unit cell with $a = 1.0226(2)$ nm, which is slightly larger than the previously reported value of 1.0219 nm.

Resistivity and specific heat measurements were carried out on many crystals in a Quantum Design PPMS system equipped with a $^3$He refrigerator down to 0.4 K. The former was measured by the standard four probe method, and the latter was measured by the heat-relaxation method using a crystal of about 10 mg in weight.

## 3. Results and discussion

We found a very sharp drop in resistivity due to superconductivity, as typically shown in the inset to Fig. 2. The onset temperature is 1.06 K and a zero-resistivity within our experimental resolution of ~ 10 nV for the voltage detection is attained below 0.99 K. Correspondingly, a large diamagnetic signal due to the Meissner effect was observed below 1.0 K [9]. The resistivity at 300 K and just above the $T_c$ are 320μΩcm and 11.5 μΩcm, respectively, which gives a residual resistivity ratio (RRR) of 28. The crystal used for the measurement is #1G which refers crystal G picked up from sample batch #1. The RRR varied from 10 to 30, depending on sample batches. The temperature dependence of resistivity at low temperature of $1$ K $< T <$ $30$ K is approximately proportional to $T^3$, not $T^2$, suggesting that electron-electron scattering is not predominant there. We have invariably observed this $T^3$ behavior, irrespective of sample batches. In contrast, Jin $et$ $al.$ reported a $T^2$ behavior below $T = 60$ K on their resistivity measurements [11]. On the other hand, the temperature dependence of resistivity at high temperatures are quite unusual. It is almost temperature independent around room temperature, while, with cooling down, it suddenly starts to decrease at about 200 K. There is a second-order structural transition at 200 K which must dramatically affect the electronic structure of $Cd_2Re_2O_7$ [15, 16].

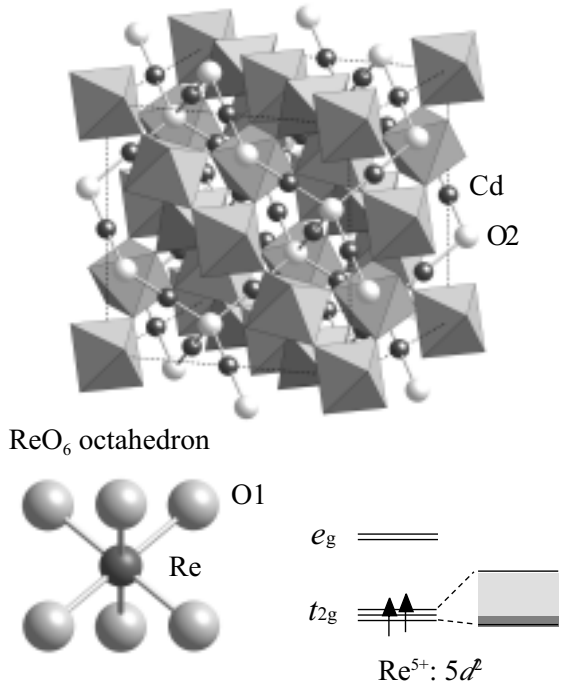

Fig. 1. Pyrochlore structure for $Cd_2Re_2O_7$ comprising $Re(O1)_6$ octahedra. Cadmium and another oxygen ions (O2) are depicted with small dark and bright balls, respectively. An electron configuration for $Re^{5+}$ with two $5d$ electrons is shown schematically.

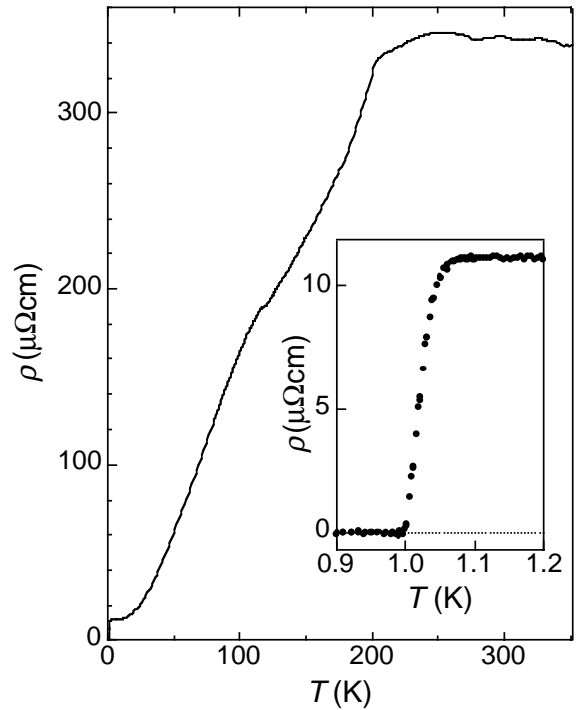

Fig. 2. Temperature dependence of resistivity measured on a $Cd_2Re_2O_7$ single crystal (#1G). The inset is an enlargement below 1.2 K showing a superconducting transition at $T_c = 1$ K.

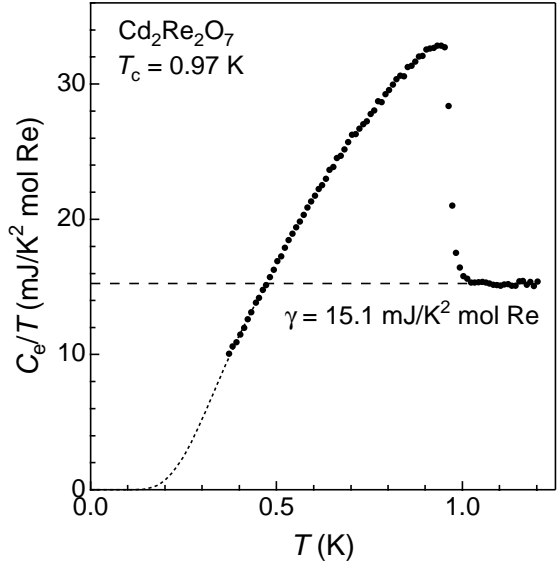

Fig. 3. Temperature dependence of electronic specific heat $C_e$ divided by temperature for a $Cd_2Re_2O_7$ single crystal (#1E) showing a superconducting transition at 0.97 K. The horizontal broken line shows the $C_e/T$ expected for the normal state, and the dotted line below 0.5 K is a fit to the form $C_e = T^{-3/2}\exp(-\Delta_0/k_BT)$ with $2\Delta_0/k_BT_c = 3.6$.

Figure 3 shows a superconducting transition at zero field detected by specific heat $C$ measurements. A distinct, large jump is seen at 1 K, which evidences the bulk nature of superconductivity in the present compound. By fitting the raw data between 1 K and 10 K to the form $C = \gamma T + \alpha T^3 + \beta T^5$, we obtained $\gamma = 15.1$ mJ/K$^2$ mol Re, $\alpha = 0.111$ mJ/K$^4$ mol Re, and $\beta = 1.35 \times 10^{-6}$ mJ/K$^6$ mol Re. This $\gamma$ value is slightly larger than that reported previously on a polycrystalline sample [14]. The Debye temperature deduced from $\alpha$ is 458 K. The electronic specific heat $C_e$ was obtained by subtracting the lattice contribution estimated above, and $C_e/T$ is plotted in Fig.3. The $T_c$ determined from the midpoint of the jump in $C_e/T$ is 0.97 K, and the transition width between 25% and 75% height of the jump is 30 mK. The magnitude of jump in $C_e$ at $T_c$ is 16.9 mJ/K mol Re, and thus $\Delta C_e/\gamma T_c$ is 1.15, which is considerably smaller than the value expected for superconductivity with an isotropic gap, 1.43. Since the data is limited above 0.37 K, it is difficult to discuss how the specific heat reaches zero as $T \to 0$. From the entropy balance, however, assuming that the normal-state specific heat is simply $\gamma T$, it is considered that the specific heat must decrease with temperature rather quickly as in an exponential form than in a power law as observed in $Sr_2RuO_4$ [17]. This suggests that the superconducting ground state of $Cd_2Re_2O_7$ is possibly nodeless unlike $Sr_2RuO_4$. By fitting the data below $T = 0.5$ K to the form $C_e = T^{-3/2}\exp(-\Delta_0/k_BT)$ under a constraint to keep the entropy balance between superconducting and normal states, we obtained the size of the superconducting gap $2\Delta_0/k_BT_c = 3.6$ which is very close to the value expected for a weak-coupling BCS superconductor.

The sample dependence of the superconducting transition of $Cd_2Re_2O_7$ is summarized in Fig. 4. Specific heat data measured on four crystals from different batches and one polycrystalline sample which were prepared in nearly the same conditions are shown in Fig. 4a. The $T_c$ of those four crystals scatters in a narrow temperature window between 0.97 K and 1.04 K, while that of the polycrystalline sample is apparently higher, $T_c = 1.2$ K at the midpoint of the transition. In addition, a narrow transition observed for the crystals contrasts strikingly with a very broad transition for the polycrystalline sample. This suggests an inhomogeneous nature for the latter and also that the $T_c$ can be raised considerably above 1 K, though the origin is not known. Anyway, it is concluded that the bulk $T_c$ of single crystals is about 1.0 K. In remarkable contrast, the $T_c$ deduced from resistivity measurements scatters very much between $T = 1$ and 2 K, as shown in Fig. 4b. We noticed that a pristine crystal gave a lower $T_c$ near 1 K, while higher $T_c$'s up to 2 K were always attained on polished crystals. Note that there are no correlations between the values of $T_c$ and residual resistivity. This high $T_c$'s must be due to surface layers, which will be reported elsewhere.

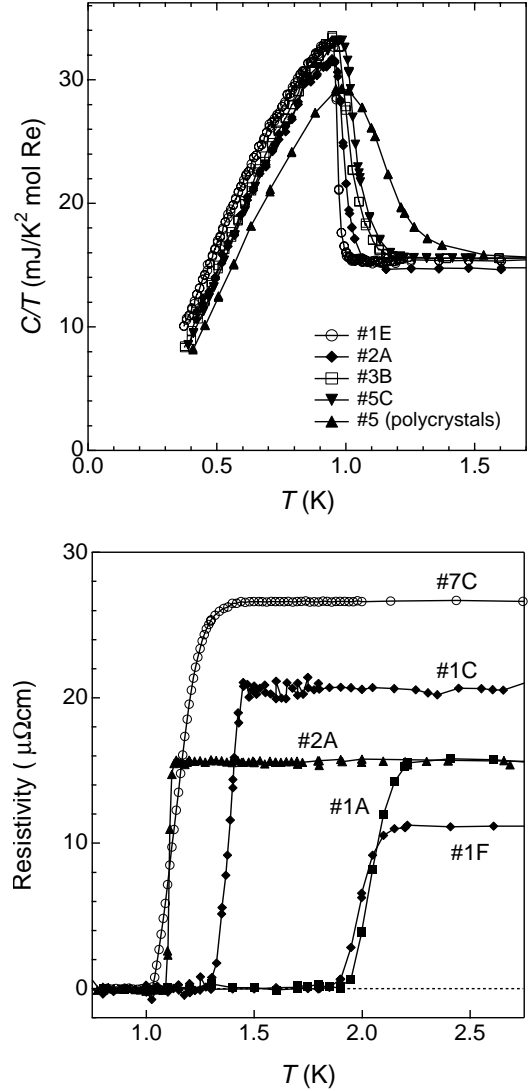

Fig. 4. Sample dependence of the superconducting transition observed by specific heat (a) and resistivity measurements (b). Data for four single crystals (#1E, #2A, #3B, and #5C) and one polycrystalline sample (#5) are shown in (a). Crystal #5C and the polycrystalline sample were prepared in the same batch. In (b) five crystals (#1A, #1C, #1F, #2A, and #7C) were used for resistivity measurements. Crystals #1A and #1F with $T_c \sim 2$ K were measured in a rectangular bar after polishing, while crystal #1C showing a lower $T_c$ was measured without any treatments. Crystals #2A and #7C with $T_c \sim 1$ K were measured after chemical etching in a diluted HCl solution.

In order to determine the upper critical field $H_{c2}$ of $Cd_2Re_2O_7$ specific heat and resistivity were measured under various magnetic fields. As shown in the $C/T$ versus $T$ plot of Fig. 5a the bulk superconductivity is completely destroyed at a rather small magnetic field of 0.5 T. The crystal used is from batch #1 with the highest quality with $\Delta T_c = 30$ mK. Considerable broadening in the transition with increasing applied field is to be noted; i.e., $\Delta T_c = 120$ mK at $\mu_0 H = 0.1$ T. The $T_c$ was determined by fitting the transition curve to that expected from the mean field theory, and its field dependence is plotted in Fig. 6. In remarkable contrast to the above results from specific heat measurements, superconductivity is much robust against field in the resistivity measurements, as shown in Fig. 5b. The crystal used there is #2A which specific heat data is shown in Fig. 4a ($T_c = 0.99$ K) and which resistivity data under zero field is shown in Fig. 4b ($T_c = 1.13$ K, 1.11 K, 1.09 K at the onset, midpoint, and offset, respectively). The measurements were done on a crystal after chemical etching in a diluted HCl solution. As seen in Fig. 5b, superconductivity survives up to $\mu_0 H = 1.2$ T at 0.4 K. Broadening of the transition such as seen in the specific heat measurements (Fig. 5a) is not discernible. The field dependence of $T_c$ which was determined as the midpoint of the transition is plotted also in Fig. 6.

Since the $H_{c2}(T)$ data obtained from the specific heat measurements is limited above $T \sim 0.5\ T_c$, it is difficult to extrapolate them to $T = 0$. Therefore, we estimated the $H_{c2}(0)$ from the initial slope $(dH_{c2}/dT)_{T=T_c}$, using the Werthamer-Helfand-Hohenberg (WHH) formula for the weak-coupling BCS theory, $H_{c2}(0) = 0.693(-dH_{c2}/dT)_{T=T_c}T_c$ [18]. It gave that $\mu_0 H_{c2}(0) = 0.29$ T. All the data from specific heat measurements and $\mu_0 H_{c2}(0)$ by the WHH formula are fitted to the form, $H_{c2}(T) = H_{c2}(0)(1-\alpha(T/T_c)^\beta)$. The result is shown by the solid line on the data with $\alpha = 0.997$ and $\beta = 1.47$. On the other hand, from the resistivity data we obtained a much larger value of 1.46 T for $\mu_0 H_{c2}(0)$ by the WHH formula. This value is close to the Pauli limiting field $\mu_0 H_P = 1.24 k_B T_c/\mu_B = 2.0$ T. Using the former value reflecting a bulk nature in the Ginzburg-Landau (GL) formula, $H_{c2}(0) = \phi_0/(2p\xi_{GL}(0)^2)$, where $\phi_0$ is the flux quantum and $\xi_{GL}(0)$ is the GL coherence length at $T = 0$, we obtained that $\xi_{GL}(0) = 34$ nm. Sakai *et al.* and Jin *et al.* reported $H_{c2}$ values of 0.8 and 0.6 T, respectively, through their resistivity measurements under magnetic fields [10, 11]. From our results, however, it is considered that these values are not for bulk superconductivity but may be attributable to superconductivity on crystal surface.

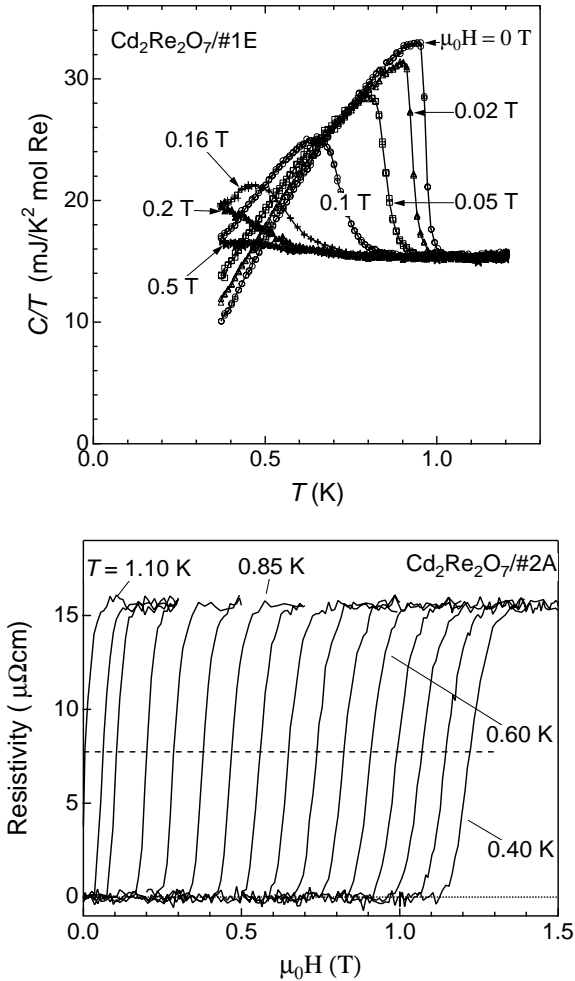

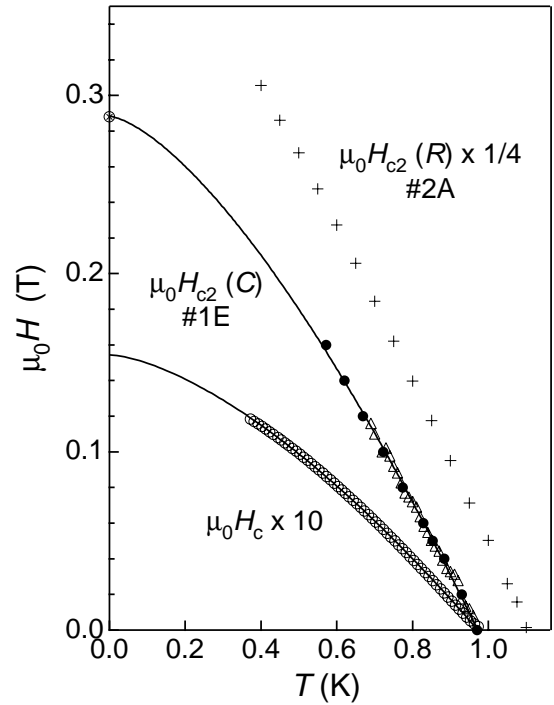

Fig. 5. Magnetic field dependence of specific heat for crystal #1E (a) and resistivity for crystal #2A (b). The $C/T$ is plotted against $T$ in (a), while the resistivity is plotted against field in (b). The curves in (b) from left to right correspond to the measurements at $T = 1.10$ K to 0.40 K with an interval of 0.05 K, respectively. The variation of $T_c$ as a function of field thus determined is plotted in Fig. 6.

Fig. 6. $H$-$T$ phase diagram showing the upper critical fields determined from the specific heat data shown in Fig. 5a (filled cricle; $\mu_0 H_{c2}(C)$) and from the resistivity data in Fig. 5b (cross; $\mu_0 H_{c2}(R)$) which is plotted after reduction by 1/4. Open triangles show $\mu_0 H_{c2}(T)$ data determined in a series of isothermal specific heat measurements as a function of field, which are not presented in this paper. The thermodynamic critical field $H_c$ is also plotted after being multiplied by 10.

Thermodynamical critical field $H_c(T)$ has been calculated from the specific heat data shown in Fig. 3, using the relation;

$$\int_T^{T_c} dT \int_T^{T_c} \frac{C_e(T') - \gamma T'}{T'} dT' = \frac{H_c(T)^2}{8\pi},$$

where $C_e$ and $\gamma T'$ are the electronic specific heat in superconducting and normal states, respectively. Thus obtained $H_c(T)$ is plotted in Fig. 6 as a function of temperature, together with a curve fitted to the form $H_c(T) = H_c(0)(1-\alpha(T/T_c)^\beta)$ with $\alpha = 0.996$, $\beta = 1.54$, and $\mu_0 H_c(0) = 0.015$ T. This $H_c(0)$ value is in good agreement with $\mu_0 H_c(0) = 0.0148$ T calculated from the BCS equation, $H_c(0)^2 = \gamma T_c^2/(0.17 m_B)$ with $\gamma = 15.1$ mJ/K$^2$ mol Re. The temperature dependence of $H_c(T)$ deviates from a simple relation of $H_c(0)(1-(T/T_c)^2)$. It also significantly deviates from what the BCS theory predicts.

Next we have calculated various superconducting parameters using the relevant theoretical equations. The results are summarized in Table 1. The GL parameter $k(0)$ is evaluated from the equation $H_{c2}(0) = \sqrt{2}\kappa(0)H_c(0)$ to be $\kappa(0) = 14$. Thus, $Cd_2Re_2O_7$ is a typical type II superconductor. The penetration depth $\lambda(0)$ is 460 nm from $\kappa(0) = \lambda(0)/\xi_{GL}(0)$. Using the relation $H_{c1}(0)H_{c2}(0) = H_c(0)^2(\ln\kappa(0) + 0.08)$ valid for $\kappa(0) \gg 1$ gives the lower critical field $H_{c1}(0)$ of 0.002 T, in good agreement with that obtained by magnetization measurements [9]. We have carried out Hall measurements which will be reported elsewhere, and obtained the Hall coefficient $R_H = -3.2 \times 10^{-10}$ m$^3$/C at temperature below 50 K. This value corresponds to carrier concentration $n$ of nearly one electron per Re in a simple one-band picture. Then, we estimated the mean-free-path $l$ to be 20 nm from the $n$ and residual resistivity $\rho_0$ of 10 μΩcm, using the equation;

$$\rho_0 = \frac{\hbar(3\pi^2)^{\frac{1}{3}}}{e^2 \ell} n^{-\frac{2}{3}}$$

However, this is obviously underestimation, because electronic structure calculations have pointed out a semi-metallic nature for the band structure of $Cd_2Re_2O_7$ [12, 13]. Assuming carrier concentration of $5 \times 10^{-3}$ per Re atom from the calculation [12] leads us to $l \sim 700$ nm. Since this $l$ value is much larger than $\xi_{GL}(0) = 34$ nm, the superconductivity of $Cd_2Re_2O_7$ must lie in the clean limit.

## 4. Concluding remarks

We have reported the superconducting properties of the pyrochlore oxide $Cd_2Re_2O_7$. The bulk $T_c$ is about 1.0 K and the upper critical field $H_{c2}(0)$ is 0.29 T. The coherence length is estimated to be 34 nm. Specific heat data measured on single crystals suggest that the superconducting ground state of $Cd_2Re_2O_7$ has a nodeless gap. Very recent [187]Re NQR measurements detected a large coherence peak below 1 K to be interpreted within the weak-coupling BCS model with a nearly isotropic energy gap [19].

We have expected that the mechanism of superconductivity in $Cd_2Re_2O_7$ is not conventional but exotic from naive considerations on the pyrochlore oxides: It occurs on the highly frustrated lattice; there exist similar compounds nearby showing a metal-insulator transition like $Cd_2Os_2O_7$; ferromagnetic spin correlations may be dominant instead of antiferromagnetic ones, as generally seen in many pyrochlore oxides. Moreover, especially for Re pyrochlores, a charge fluctuation might be present [1], because $Re^{5+}$ is believed to be unstable compared with $Re^{6+}$ or $Re^{4+}$ in oxide system [20]. However, we have little experimental evidence to imply exotic superconductivity at the moment. To be noted is that Kadono measured the penetration depth using the mSR technique on our polycrystalline and single-crystal samples and found an unconventional temperature dependence suggesting an anisotropic superconducting gap [21]. Further information is necessary to discuss on the mechanism.

## Acknowledgements

We would like to thank F. Sakai for chemical analysis and T. Yamauchi for valuable discussion. This research was supported by a Grant-in-Aid for Scientific Research on Priority Areas (A) and a Grant-in-Aid for Creative Scientific Research given by The Ministry of Education, Culture, Sports, Science and Technology, Japan.